%% file: main.tex
\documentclass[copyright,creativecommons]{eptcs}
\providecommand{\event}{PLACES 2016}

\usepackage{hyperref} 

\input{def/fonts}
\input{def/packages}

\input{def/layout}
\input{def/listings}
\input{def/macros}

\let\citet=\cite

\newcommand*{\Title}{Multiparty Compatibility for Concurrent Objects}
\input{def/authors/eptcs}

\begin{document}
\title{\Title}
\maketitle

\input{abstract}
\input{sec/introduction}
\input{sec/methodology}
\input{sec/theory}
\input{sec/conclusion}
\input{acknowledgements}

\input{main.bbl}
\bibliographystyle{eptcs}

\end{document}

%% file: def/fonts.tex
\usepackage[T1]{fontenc}
\usepackage[scaled=0.75]{beramono}
\usepackage{upgreek}

%% file: def/packages.tex
\usepackage{extpfeil}
\usepackage{amsmath}
\usepackage{amsthm}
\usepackage{amssymb}
\usepackage{float}
\usepackage{stmaryrd}
\usepackage[usenames,dvipsnames]{color}
\usepackage{mathpartir}
\usepackage{microtype}
\usepackage{tikz}
\usepackage{xspace}
\usepackage{calc} 
\usepackage{pifont} 
\usepackage{listings}

\usepackage{bm}
\usetikzlibrary{automata}
\usetikzlibrary{positioning}
\usepackage{mathtools}
\usepackage{cancel}
\usepackage{array}
\usepackage{tikz-cd}
\usepackage[strings]{underscore}
\usepackage{ulem}
\input{def-import/macros}


\input{def-import/MnSymbol-twoheadarrows}
\input{def-import/MnSymbol-powerset}
\input{def-import/MnSymbol-triangles}

%% file: def-import/macros.tex
\newcommand*{\ttt}[1]{\texttt{#1}}
\newcommand*{\kw}[1]{{\text{\tt{#1}}}} 

\DeclareSymbolFont{bbsymbol}{U}{bbold}{m}{n}
\DeclareMathSymbol{\bbsemi}{\mathbin}{bbsymbol}{"3B}

\newcommand*{\figref}[1]{Figure~\ref{fig:#1}}

\newcommand*{\secref}[1]{Section~\ref{sec:#1}}

\newcommand*{\Secref}[1]{\S\,\ref{sec:#1}}

\newenvironment{nop}{}{}
\newenvironment{sdisplaymath}{
\begin{nop}\small\begin{displaymath}}{
\end{displaymath}\end{nop}\ignorespacesafterend}

\newenvironment{mathfig}{\begin{sdisplaymath}}{\end{sdisplaymath}}

\makeatletter
\newbox\sf@box
  {\def\sf@one{#1}%
   \def\sf@two{#2}%
   \setbox\sf@box\hbox
     \bgroup}%
  { \egroup
   \ifx\@empty\sf@two\@empty\relax
     \def\sf@two{\@empty}
   \fi
   \ifx\@empty\sf@one\@empty\relax
     \subfloat[\sf@two]{\box\sf@box}%
   \else
     \subfloat[\sf@one][\sf@two]{\box\sf@box}%
   \fi}
\makeatother

\definecolor{highlightcolor}{rgb}{1.0,0.8,0.8}
\definecolor{shadecolor}{rgb}{0.9,0.9,0.9}
\definecolor{lightgray}{rgb}{0.8,0.8,0.8}

\newenvironment{nscenter}
 {\parskip=0pt\par\nopagebreak\centering}
 {\par\noindent\ignorespacesafterend}

\newsavebox{\vardisplaymathbox}

%% file: def-import/MnSymbol-twoheadarrows.tex
\DeclareFontFamily{U}{MnSymbolB}{}
\DeclareFontShape{U}{MnSymbolB}{m}{n}{
    <-6>  MnSymbolB5
   <6-7>  MnSymbolB6
   <7-8>  MnSymbolB7
   <8-9>  MnSymbolB8
   <9-10> MnSymbolB9
  <10-12> MnSymbolB10
  <12->   MnSymbolB12}{}
\DeclareFontShape{U}{MnSymbolB}{b}{n}{
    <-6>  MnSymbolB-Bold5
   <6-7>  MnSymbolB-Bold6
   <7-8>  MnSymbolB-Bold7
   <8-9>  MnSymbolB-Bold8
   <9-10> MnSymbolB-Bold9
  <10-12> MnSymbolB-Bold10
  <12->   MnSymbolB-Bold12}{}
\DeclareSymbolFont{MnSyB}{U}{MnSymbolB}{m}{n}

\DeclareFontFamily{U}{MnSymbolA}{}
\DeclareFontShape{U}{MnSymbolA}{m}{n}{
    <-6>  MnSymbolA5
   <6-7>  MnSymbolA6
   <7-8>  MnSymbolA7
   <8-9>  MnSymbolA8
   <9-10> MnSymbolA9
  <10-12> MnSymbolA10
  <12->   MnSymbolA12}{}
\DeclareFontShape{U}{MnSymbolA}{b}{n}{
    <-6>  MnSymbolA-Aold5
   <6-7>  MnSymbolA-Aold6
   <7-8>  MnSymbolA-Aold7
   <8-9>  MnSymbolA-Aold8
   <9-10> MnSymbolA-Aold9
  <10-12> MnSymbolA-Aold10
  <12->   MnSymbolA-Aold12}{}
\DeclareSymbolFont{MnSyA}{U}{MnSymbolA}{m}{n}

\let\twoheaddownarrow\relax
\DeclareMathSymbol{\twoheaddownarrow}{\mathrel}{MnSyA}{27}
\let\twoheaduparrow\relax
\DeclareMathSymbol{\twoheaduparrow}{\mathrel}{MnSyA}{25}

\let\ntwoheaddownarrow\relax
\DeclareMathSymbol{\ntwoheaddownarrow}{\mathrel}{MnSyB}{27}
\let\ntwoheaduparrow\relax
\DeclareMathSymbol{\ntwoheaduparrow}{\mathrel}{MnSyB}{25}

%% file: def-import/MnSymbol-powerset.tex
\DeclareFontFamily{U}{MnSymbolC}{}
\DeclareFontShape{U}{MnSymbolC}{m}{n}{
    <-6>  MnSymbolC5
   <6-7>  MnSymbolC6
   <7-8>  MnSymbolC7
   <8-9>  MnSymbolC8
   <9-10> MnSymbolC9
  <10-12> MnSymbolC10
  <12->   MnSymbolC12}{}
\DeclareFontShape{U}{MnSymbolC}{b}{n}{
    <-6>  MnSymbolC-Bold5
   <6-7>  MnSymbolC-Bold6
   <7-8>  MnSymbolC-Bold7
   <8-9>  MnSymbolC-Bold8
   <9-10> MnSymbolC-Bold9
  <10-12> MnSymbolC-Bold10
  <12->   MnSymbolC-Bold12}{}
\DeclareSymbolFont{MnSyC}{U}{MnSymbolC}{m}{n}

\DeclareMathSymbol{\MnSymbolpowerset}{\mathord}{MnSyC}{180}

%% file: def-import/MnSymbol-triangles.tex
\DeclareFontFamily{U}{MnSymbolC}{}
\DeclareFontShape{U}{MnSymbolC}{m}{n}{
    <-6>  MnSymbolC5
   <6-7>  MnSymbolC6
   <7-8>  MnSymbolC7
   <8-9>  MnSymbolC8
   <9-10> MnSymbolC9
  <10-12> MnSymbolC10
  <12->   MnSymbolC12}{}
\DeclareFontShape{U}{MnSymbolC}{b}{n}{
    <-6>  MnSymbolC-Bold5
   <6-7>  MnSymbolC-Bold6
   <7-8>  MnSymbolC-Bold7
   <8-9>  MnSymbolC-Bold8
   <9-10> MnSymbolC-Bold9
  <10-12> MnSymbolC-Bold10
  <12->   MnSymbolC-Bold12}{}
\DeclareSymbolFont{MnSyC}{U}{MnSymbolC}{m}{n}

\DeclareMathSymbol{\smalltriangleright}{\mathord}{MnSyC}{72}
\DeclareMathSymbol{\smalltriangleup}{\mathord}{MnSyC}{73}
\DeclareMathSymbol{\smalltriangleleft}{\mathord}{MnSyC}{74}
\DeclareMathSymbol{\smalltriangledown}{\mathord}{MnSyC}{75}
\DeclareMathSymbol{\filledtriangleright}{\mathord}{MnSyC}{76}
\DeclareMathSymbol{\filledtriangleup}{\mathord}{MnSyC}{77}
\DeclareMathSymbol{\filledtriangleleft}{\mathord}{MnSyC}{78}
\DeclareMathSymbol{\filledtriangledown}{\mathord}{MnSyC}{79}

%% file: def/layout.tex
\everymath{\displaystyle}

\allowdisplaybreaks

\usepackage{microtype}

\hypersetup{ 
    colorlinks=false,
    pdfborder={0 0 0},
}

%% file: def/listings.tex
\definecolor{verylightgray}{gray}{0.9}
\definecolor{lightgray}{gray}{0.5}
\definecolor{mediumgray}{gray}{0.45}
\newlength\lsthorizontalpadding
\newcommand*\lstnumberstyle{\ttfamily\scriptsize\textcolor{lightgray}}
\newlength\lstnumbersep
\newlength\lstnumberwidth
\lstdefinelanguage{code}{%
   morekeywords={system,behaviour,obj,using}%
  ,morecomment=[s]{\{-}{-\}}%
}
\lstnewenvironment{codecol}[1][]{\lstset{language=code,#1,moredelim=*[is][\textcolor{mediumgray}]{|}{|}}}{}

%% file: def/macros.tex
\newcommand*{\channel}[2]{\actorId{#1}\actorId{#2}}

\newcommand*{\actorId}[1]{\mathsf{#1}}

\newcommand*{\Receive}[1]{?{#1}}
\newcommand*{\Send}[1]{!{#1}}
\newcommand*{\singleton}[3]{\channel{#1}{#2}{#3}}

\newcommand*{\tensor}{\otimes}

\newcommand*{\Compatible}{\Join}

\newcommand*{\subbeh}{\lesssim}

\newcommand*{\sendTriangle}{\filledtriangleleft}
\newcommand*{\receiveTriangle}{\filledtriangleright}

\makeatletter
\def\redwave{\bgroup \markoverwith{\lower3\p@\hbox{\sixly \textcolor{red}{\char58}}}\ULon}
\def\bluewave{\bgroup \markoverwith{\lower3\p@\hbox{\sixly \textcolor{blue}{\char58}}}\ULon}
\font\sixly=lasy6 
\makeatother
\newcommand\reduline{\bgroup\markoverwith{\textcolor{red}{\rule[-0.5ex]{2pt}{0.4pt}}}\ULon}
\newcommand\blueuline{\bgroup\markoverwith{\textcolor{blue}{\rule[-0.5ex]{2pt}{0.4pt}}}\ULon}

\newcommand*{\inscolour}{OliveGreen}

\newcommand*{\external}[1]{\smash{\texttt{\emph{#1}}}}

\newcommand*{\ins}[1]{\smash{\texttt{\textcolor{\inscolour}{#1}}}}
\newcommand*{\insmath}[1]{\textcolor{\inscolour}{#1}}
\newcommand*{\receiveerr}[1]{\smash{\texttt{\bluewave{#1}}}}
\newcommand*{\senderr}[1]{\smash{\texttt{\redwave{#1}}}}
\newcommand*{\receivebad}[1]{\smash{\texttt{\blueuline{#1}}}}
\newcommand*{\sendbad}[1]{\smash{\texttt{\reduline{#1}}}}

\renewcommand*{\secref}{\Secref}

\tikzstyle{negative} = [circle, minimum width=8pt, fill, inner sep=0pt]
\tikzstyle{positive} = [circle, minimum width=8pt, draw, inner sep=0pt]
\tikzstyle{mixed} = [circle, minimum width=8pt, draw=black, fill=lightgray, inner sep=0pt]

\newcommand*{\strCompatibility}{Compatible usage}
\newcommand*{\strCompliance}{Compliant refinement}
\newcommand*{\strTestOrientedDevelopment}{Test-oriented methodologies}

%% file: def/authors/eptcs.tex
\author{
  Roly Perera
  \institute{University of Glasgow, UK}
  \institute{School of Computing Science}
  \email{roly.perera@glasgow.ac.uk}
  \and
  Julien Lange
  \institute{Imperial College London}
  \institute{Department of Computing}
  \email{j.lange@imperial.ac.uk}
  \and
  Simon J. Gay
  \institute{University of Glasgow, UK}
  \institute{School of Computing Science}
  \email{simon.gay@glasgow.ac.uk}
}
\def\titlerunning{\Title}
\def\authorrunning{R. Perera, J. Lange \& S. J. Gay}

%% file: abstract.tex
\begin{abstract}
Objects and actors are communicating state machines, offering and
consuming different services at different points in their lifecycle. Two
complementary challenges arise when programming such systems. When
objects interact, their state machines must be ``compatible'', so that
services are requested only when they are available. Dually, when
objects refine other objects, their state machines must be
``compliant'', so that services are honoured whenever they are promised.

In this paper we show how the idea of \emph{multiparty compatibility}
from the session types literature can be applied to both of these
problems. We present an untyped language in which concurrent objects are
checked automatically for compatibility and compliance. For simple
objects, checking can be exhaustive and has the feel of a type system.
More complex objects can be partially validated via test cases, leading
to a methodology closer to continuous testing. Our proof-of-concept
implementation is limited in some important respects, but demonstrates
the potential value of the approach and the relationship to existing
software development practices.
\end{abstract}

%% file: sec/introduction.tex
\section{Objects as communicating automata}
\label{sec:introduction}

Two significant state-related challenges arise when programming
object-oriented systems. We describe how these both relate to the idea
of an object as an automaton, and then present a tool which speaks to
both challenges and aligns well with test-driven development practices.

\paragraph{\strCompatibility.}
To use an object safely, one must understand its inner automaton: its states,
the services available in each state, and the state transitions that
result from service requests. We call this the \emph{compatibility
  problem}. In standard OO languages the automaton is hidden away behind
a flat collection of methods comprising the object's interface. The
compatibility problem is compounded by the fact that interacting with
one object may indirectly cause a state change in another object.

\paragraph{\strCompliance.}
The second challenge is complementary. To safely specialise the
behaviour of an object, one must take care to respect its automaton:
each state of the overriding object must be contravariant with respect
to services offered, and covariant with respect to services consumed. We
call this the \emph{compliance problem}. Compliance is the key to robust
compositionality: compliant implementations can be safely substituted
for abstractions, as embodied by the slogan ``require no more, promise
no less'' \cite{liskov87}, allowing the abstractions to serve as
boundaries between components. Yet in traditional OO languages
compliance is captured at the level of method signatures, not at the
level of the object's underlying automaton.

\paragraph{\strTestOrientedDevelopment.}
The modern context for both of these problems is an increasingly
incremental and test-driven development process. In test-driven
development \cite{beck02}, programmers deliver features by first
defining test cases which characterise them, and then writing enough
code to make the tests pass. Having a suite of tests makes it easier to
refactor code and increases the visibility of changes to observable
behaviour.

Test-driven development has evolved into a host of sophisticated
techniques and tools. \emph{Mock objects} \cite{freeman09} are a way of
simulating, in the test environment, the other participants in a
multi-agent system. \emph{Continuous testing} \cite{saff04,madeyski13}
is a tool feature that automatically runs tests in the background to
shorten the feedback cycle between code changes and test failures. These
and similar practices are the context in which we wish to propose a new
approach to the compatibility and compliance problems.

\subsection{Contributions of paper}

Our proposal is an approach to language implementation which integrates
compatibility checking and compliance checking of automata directly into
the programming environment (\secref{methodology}). This somewhat blurs
the distinction between language and tool: an interactive, incremental
style of programming, with immediate feedback on failures, is baked into
the way the language works. The distinguishing feature of our approach is
that there are no types: compatibility and compliance checking is at the
level of objects, making the approach a form of language-integrated
continuous testing. Any object or system of objects may serve both as an
abstraction and as a prototypical implementation.

\input{fig/example/release-cycle/dev-refactored}

Our goal in this paper is to motivate our language/tool design and the
development methodology it supports (\secref{methodology}). In
\secref{theory} we outline the theoretical developments that will be
needed to put the approach on a sound footing, including possible
extensions and modifications to the notions of communicating automata
\cite{brand83} and multiparty compatibility
\cite{denielou12,denielou13,bocchi14,carbone15,lange15} from the theory
of session types. We also mention connections to established techniques
such as model checking.

%% file: fig/example/release-cycle/dev-refactored.tex
\begin{figure*}[ht]
\begin{nscenter}
\begin{minipage}[t]{0.45\textwidth}
\small
\begin{lstlisting}
system dev

obj teamLead
behaviour ReleaseCycle
   devTeam$\receiveTriangle$releaseCandidate
   $\external{business}$$\sendTriangle$evaluate
   $\external{business}$$\receiveTriangle${
      $\receivebad{iterate}$(tag)
         $\external{repository}$$\sendTriangle$tagRC(tag)
         devTeam$\sendTriangle$continue
         ReleaseCycle
      accept(tag)
         $\external{repository}$$\sendTriangle$tagRelease(tag)
         devTeam$\sendTriangle\senderr{stop}$.
   }
ReleaseCycle

obj devTeam
$\external{repository}$$\sendTriangle$commit
$\external{repository}$$\receiveTriangle$revision(n)
behaviour ReleaseCycle
   teamLead$\sendTriangle$releaseCandidate
   teamLead$\receiveTriangle${
      $\receiveerr{continue}$
         $\external{repository}$$\sendTriangle$commit
         $\external{repository}$$\receiveTriangle$revision(n)
         ReleaseCycle
   }
ReleaseCycle
\end{lstlisting}
\end{minipage}%
\quad\quad\quad%
\begin{minipage}[t]{0.45\textwidth}
\small
\begin{lstlisting}
system dev-refactored: dev

obj teamLead
behaviour ReleaseCycle
   devTeam$\receiveTriangle$releaseCandidate
   $\external{business}$$\sendTriangle$evaluate
   $\external{business}$$\receiveTriangle${
      accept(tag)
         $\external{repository}$$\sendTriangle$$\sendbad{tagRC}$(tag)
         devTeam$\sendTriangle$stop.
   }
ReleaseCycle

obj devTeam
behaviour ReleaseCycle
   $\external{repository}$$\sendTriangle$commit
   $\external{repository}$$\receiveTriangle$revision(n)
   teamLead$\sendTriangle$releaseCandidate
   teamLead$\receiveTriangle${
      continue
         ReleaseCycle
      stop.
   }
ReleaseCycle
\end{lstlisting}
\end{minipage}
\end{nscenter}
\caption{Partially refactored \ttt{dev} system, with two implementation errors}
\label{fig:example:release-cycle:dev-refactored}
\end{figure*}

%% file: sec/methodology.tex
\section{Language-integrated verification}
\label{sec:methodology}

\emph{Actors} \cite{agha86,hewitt73} are the natural setting for
exploring our proposed tool design, since we are concerned with an
object's ``inner automaton''. Unlike other OO paradigms actors make this
automaton explicit, allowing the object to change its interface at
runtime.

We start with a fairly standard actor calculus and then generalise in
some respects and restrict in others. An asynchronous send to an actor
\kw{p} is non-standard in allowing one of a set of messages to be chosen
non-deterministically, and is written
\kw{p$\sendTriangle$\{m$_{\text{1}}$(v$_{\text{1}}$)P$_{\text{1}}$\ldots
  m$_{\text{n}}$(v$_{\text{n}}$)P$_{\text{n}}$\}}. Non-determinism
allows objects to be general enough to serve as specifications, but is
also convenient for testing. A blocking receive which waits until one of
a set of possible messages arrives from \kw{p} is written
\kw{p$\receiveTriangle$\{m$_{\text{1}}$(x$_{\text{1}}$)P$_{\text{1}}$\ldots
  m$_{\text{n}}$(x$_{\text{n}}$)P$_{\text{n}}$\}}, and binds the
parameter \kw{x} to the value received. An unrealistically strong but
expedient assumption for now is that objects cannot be allocated
dynamically but have a fixed configuration. Finally, an actor maintains
a unidirectional FIFO queue per client object.

Our implementation is only intended as a proof-of-concept and as such is
unable to produce standalone code. A hosted version will be made
available online before the workshop.

\paragraph{\strCompatibility.}
Our tool's support for compatibility testing is illustrated on the
left-hand side of \figref{example:release-cycle:dev-refactored}. The
code models a simple software development workflow with four objects, of
which only two, \ttt{teamLead} and \ttt{devTeam}, are defined here.

The team lead iterates through a release cycle. At each cycle, the team
lead waits to receive a release candidate from the dev team, which the
business must then evaluate. The business responds either by instructing
the team lead to iterate again, or by accepting the release. In the
first case, the team lead tags the release candidate in the repository
and repeats, having told the dev team to continue. Otherwise, the
release candidate is tagged as an official release in the repository and
the dev team instructed to stop. The dev team has its own notion of
release cycle; on each iteration it commits code to the repository,
notifies the team lead, and then awaits an instruction to continue.

Our tool highlights two compatibility errors using wavy underlining. It
has detected a state in which the message \ttt{stop} message sent to the
dev team will never be delivered, and another state in which the
\ttt{continue} message that the team lead is waiting for will never
arrive. (The colour of the underlining reflects the polarity of the
message -- either sending or receiving -- but is technically redundant
given the arrows on the messages.) An important detail to notice here
is that this system is not closed: the objects \ttt{\emph{business}} and
\ttt{\emph{repository}} are unspecified external agents (indicated by
the italicised names).

\paragraph{\strCompliance.}
Compliance testing is also illustrated in
\figref{example:release-cycle:dev-refactored}. On the right-hand side of
the figure is \ttt{dev-refactored}, an attempt to refine the observable
behaviour of \ttt{dev}. The programmer specifies this intention using
the colon syntax in the \ttt{system} definition. The colon is analogous
to Java \ttt{implements} rather than Java \ttt{extends}, as no
implementation is inherited from \ttt{dev}.

Our tool highlights two compliance errors using solid underlining.
First, the programmer has not implemented \ttt{iterate}. If
\ttt{dev-refactored} were substituted for \ttt{dev}, it would be unable
to satisfy its contract. This is analogous to failing to implement a
Java interface method, except that in this language the interface (set
of available methods) varies explicitly from state to state. Although
the system which is at fault is \ttt{dev-refactored}, the error is
indicated by underlining the unmet obligation in \ttt{dev}, namely the
\ttt{iterate} handler. (The errors should be understood as a view
contextualised to \ttt{dev-refactored}.)

The second error is in the new implementation of \ttt{accept}, where the
programmer has called \ttt{tagRC} rather than \ttt{tagRelease}. If
\ttt{dev-refactored} were substituted for \ttt{dev}, it would require
services of its environment that might not be on offer. This error has
no analogue in Java's type system, since it corresponds to checking that
the \emph{implementation} of an overriding method refines the observable
behaviour of the overridden method.

Moreover, the programmer was able to consolidate the duplicate
\ttt{commit}/\ttt{revision} interaction with the repository in the
original dev team code into a single interaction at the beginning of the
loop in the new code. Our implementation is able to verify that this is
a behaviour-preserving change. This emphasises how in our language,
behavioural specifications are \emph{prototypical implementations} that
can form part of an actual system, or be refined into more specialised
implementations. The design pattern \emph{template method}
\cite{gamma95} is similar in intent but requires the programmer to
decide in advance which aspects of the behaviour are fixed and which can
be overridden by a refining implementation.

\paragraph{\strTestOrientedDevelopment.}
The compatibility and compliance features just described are related to
\emph{continuous testing} \cite{saff04,madeyski13}, in that testing is
automatic and feedback is immediate. However our approach provides more
automation than continuous testing tools, which automate test execution
but still require the programmer to write the tests in the first place.
In our implementation all possible interactions between simple
finite-state objects are verified automatically without the programmer
having to write explicit tests.

\input{fig/example/release-cycle/business}

Of course exhaustively verifying even a finite-state system quickly
becomes impractical as its state-space grows. While techniques from
model checking \cite{emerson80,queille82,lange16} and symbolic execution
\cite{boyer75} may be applicable, this problem can also be addressed by
adopting a more manual test-oriented style. The system to be tested is
placed into an environment which has been simplified by replacing some
components by \emph{mock objects} \cite{freeman09} so that the
state-space of the resulting system is small enough for exhaustive
checking to be feasible.

This is illustrated in \figref{example:release-cycle:business} on the
previous page. The repository is the code to be tested; its states are
indexed with a natural number incremented on every \ttt{commit},
representing the current revision; there is therefore a state for every
natural number. The verification of this object is made tractable by
composing it with a simplified environment: in this case a mock business
object which hard-codes a representative sequence of interactions.
(Other components are imported from \ttt{dev} by the \ttt{using}
declaration.) The resulting system has a small number of states which
our tool can automatically and exhaustively verify. Like testing in
general, this method is complete (all reported errors are \emph{bona
  fide} errors) but not sound with respect to the larger system being
tested, which may contain problems not revealed by the test.

\input{fig/example/release-cycle/discard}

Notice how non-deterministic choice substantially reduces the coding
overhead usually associated with testing. In
\figref{example:release-cycle:discard} the programmer implements a new
service on the team lead, giving the business the option to discard the
release candidate (left, green highlight); if this happens the
repository is reverted to the previous revision, and the dev team told
to start a new iteration. Rather than writing a whole new test case, the
programmer can simply add a \ttt{discard} branch to the existing test
case (right, green highlight). This immediately yields errors on
\ttt{tagRC} and \ttt{tagRelease}, and the complementary error on
\ttt{revert}, reflecting the pending obligation to add \ttt{revert} as a
service to the repository.

%% file: fig/example/release-cycle/business.tex
\begin{figure}[H]
\begin{nscenter}
\begin{minipage}[t]{0.45\textwidth}
\small
\begin{lstlisting}
system repo

obj repository
behaviour Connected(n)
   devTeam$\receiveTriangle$commit
   devTeam$\sendTriangle$revision(n)
   teamLead$\receiveTriangle${
      tagRC(tag)
         $\external{math}$$\sendTriangle$plus(n, 1)
         $\external{math}$$\receiveTriangle$val(m)
         Connected(m)
      tagRelease(tag).
   }
Connected(0)
\end{lstlisting}
\end{minipage}%
\quad\quad\quad%
\begin{minipage}[t]{0.45\textwidth}
\small
\begin{lstlisting}
system repo-test
using repo
using dev

obj business
teamLead$\receiveTriangle$evaluate
teamLead$\sendTriangle${
   accept("1.0").
   iterate("1.0RC")
      teamLead$\receiveTriangle$evaluate
      teamLead$\sendTriangle$accept("1.0").
}
\end{lstlisting}
\end{minipage}
\end{nscenter}
\caption{Non-finite state \ttt{repository}, with \ttt{business} test case}
\label{fig:example:release-cycle:business}
\end{figure}

%% file: fig/example/release-cycle/discard.tex
\begin{figure*}[ht]
\begin{nscenter}
\begin{minipage}[t]{0.45\textwidth}
\small
\begin{lstlisting}
obj teamLead
behaviour ReleaseCycle
   $\receiveerr{devTeam}$$\receiveTriangle$releaseCandidate
   business$\sendTriangle$evaluate
   business$\receiveTriangle${
      iterate(tag)
         $\external{repository}$$\sendTriangle$tagRC(tag)
         devTeam$\sendTriangle$continue
         ReleaseCycle
      accept(tag)
         $\external{repository}$$\sendTriangle$tagRelease(tag)
         devTeam$\sendTriangle$stop.
      $\ins{discard}$
         $\ins{\emph{repository}}$$\insmath{\sendTriangle}$$\senderr{\ins{revert}}$
         $\ins{devTeam}$$\insmath{\sendTriangle}$$\ins{continue}$
         $\ins{ReleaseCycle}$
   }
ReleaseCycle
\end{lstlisting}
\end{minipage}%
\quad\quad\quad%
\begin{minipage}[t]{0.45\textwidth}
\small
\begin{lstlisting}
obj repository
behaviour Connected(n)
   devTeam$\receiveTriangle$commit
   devTeam$\sendTriangle$revision(n)
   teamLead$\receiveTriangle${
      $\receiveerr{tagRC}$(tag)
         math$\sendTriangle$plus(n, 1)
         math$\receiveTriangle$val(m)
         Connected(m)
      $\receiveerr{tagRelease}$(tag).
   }
Connected(0)

obj business
teamLead$\receiveTriangle$evaluate
teamLead$\sendTriangle${
   accept("1.0").
   iterate("1.0RC")
      teamLead$\receiveTriangle$evaluate
      teamLead$\sendTriangle$$\ins{\{}$
         accept("1.0").
         $\ins{discard.}$
      $\ins{\}}$
}
\end{lstlisting}
\end{minipage}
\end{nscenter}
\caption{New \ttt{discard} service (left) verified by adding branch to test case (centre)}
\label{fig:example:release-cycle:discard}
\end{figure*}

%% file: sec/theory.tex
\section{Extensions to communicating automata and multiparty compatibility}
\label{sec:theory}

Several extensions to the theory of communicating automata and
multiparty compatibility will be required to formalise our proposed
approach. We outline some of the basic ideas here, assuming some
familarity with the usual definitions. The extensions we anticipate are
compositional multiparty compatibility
(\secref{theory:compositional-mc}), dynamic allocation of objects
(\secref{theory:dynamic-allocation}) and communicating automata that
exchange values (\secref{theory:values}).

\subsection{Compositional multiparty compatibility}
\label{sec:theory:compositional-mc}

The first requirement is to extend multiparty compatibility to systems
of objects where not all the actors are known. Here we sketch the
approach used in our implementation. First we define a notion of
composition of automata that distinguishes between ``internal'' and
``external'' transitions, illustrated here by example. In
\figref{automata:composition} the basic automata $C$ and $D$ have
$\actorId{p}$ and $\actorId{q}$ respectively as their subjects. A
message between $\actorId{p}$ and $\actorId{q}$ is therefore
\emph{internal} to the composite $C \tensor D$, since both participants
belong to the composite automaton. A message is \emph{external} iff it
is not internal.

\input{fig/automata/composition}

For brevity internal actions are shown as $\tau$ in the figure but
formally they are synchronous transitions of the form
$\actorId{p}\rightarrow\actorId{q}:i(v)$. In the context of $C \tensor
D$ it is then meaningful to ask whether $C$ and $D$ are
\emph{compatible}, written $C \Compatible D$. For this we use a
definition of multiparty compatibility in the style of \citet{bocchi15},
except that we only require that \emph{internal} interactions of $C$ and
$D$ always have a potential rendezvous partner. In
\figref{automata:composition} the automata are compatible; the initial
states of $C$ and $D$ are the only ones that need checking because the
others involve only external transitions.

In \figref{automata:composition-incompatible}, we compose the composite
automaton $C \tensor D$ with another basic automaton $C'$, which has
$\actorId{r}$ as its subject.

\input{fig/automata/composition-incompatible}

\noindent Now the internal transitions that need verifying are those
between $\actorId{p}$ and $\actorId{r}$ or between $\actorId{q}$ and
$\actorId{r}$. Assuming $k \neq m$, the transitions highlighted in red
are unable to synchronise and so are dropped from the composite $C
\tensor D \tensor C'$. A send which cannot synchronise (or dually, a set
of receives, none of which can synchronise) does not \emph{ipso facto}
represent an incompatibility, since it may synchronise in a later state.
However in this case the message $\singleton{q}{r}{\Send{k}{}}$ can
never be delivered, and so $C \tensor D$ and $C'$ are incompatible. This
incompatibility is preserved by any inclusion of $C \tensor D \tensor
C'$ into a larger system.

\subsubsection{Safety properties of compatibility}

There are two key safety properties that compatibility must guarantee.
The first is a variant of the usual property that executing $C$ and $D$
in parallel under an asynchronous semantics never leads to a deadlocked
or orphan-message configuration. The variation we need is that the
asynchronous semantics must be defined for systems where are not all
parties are present in the configuration. As with $\tensor$, we
distinguish between internal and external transitions, and then use this
to define an asynchronous semantics which only uses queues for internal
transitions. For example for \figref{automata:composition}, the
configuration would contain queues only for the channels
$\channel{p}{q}$ and $\channel{q}{p}$. External sends disappear into the
ether, and external receives never block. How this might work when
messages includes values is discussed in \secref{theory:values} below.

Even if $C$ and $D$ are compatible, it is always possible to compose
them with other objects that cause the whole system to have bad
reachable configurations. But there should be an important completeness
property, namely that if $C$ and $D$ are \emph{incompatible} then there
are no objects they can be composed with which will render them
compatible. Then any incompatibility or non-compliance errors reported
to the programmer are genuine, supporting the incremental programming
methodology we outlined in \secref{methodology}.

The second property we require for compatibility is that it be preserved
by refinement. In other words, if $C \Compatible D$ and $C' \subbeh C$
and $D' \subbeh D$ then $C' \Compatible D'$ where $\subbeh$ is the
refinement preorder (to be defined). In the more test-oriented examples
of \secref{methodology}, the objects being composed do not represent the
full set of system behaviours but \emph{test cases} that only exercise
some region of the full state-space. The property that refinement
preserves compatibility means that any properties validated by a test
case are valid for any refinements of those objects, although there will
in general be properties of the full system which are not validated by
those particular tests.

\subsection{Dynamic allocation of objects}
\label{sec:theory:dynamic-allocation}

Although the toy examples presented earlier did not require it, dynamic
object creation is essential for real applications. \citet{bollig13}
extend communicating automata with object creation; we will need to
extend multiparty compatibility to such systems.

\subsection{Communicating automata that exchange values}
\label{sec:theory:values}

A fundamental extension of the theories based on multiparty
compatibility is required to allow automata to exchange values.
The major difficulty in dealing with values is that the state-space of
the systems considered may grow exponentially, or worse, for example if
an automaton may send unbounded integers.
However, we can rely on several existing theories and techniques to
ensure enough expressivity at the automata level while keeping the logic
dealing with values decidable and manageable.
One direction to consider is translation from our automata to processes
in a modelling language such as mCRL2~\cite{groote14} which permits the
usage of sophisticated data types and associated predicates.
These can then be checked via the corresponding tool
chain~\cite{cranen13}.

On the one hand, we know that the infinite nature of communication
through unbounded FIFO channels can be dealt with through multiparty
compatibility. On the other hand, potentially infinite data types can be
dealt with through model checking theories.
%
%
However, exchanged values and pure interactions are not always
orthogonal. To deal with the effect of data on the communication
patterns, we will adopt a conservative strategy such as the one used
in~\citet{bocchi10}.


%% file: fig/automata/composition.tex
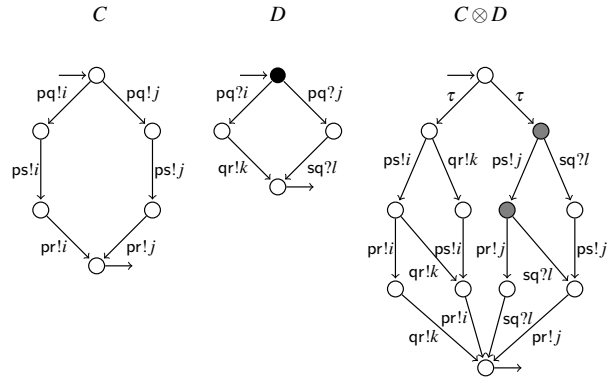
\begin{figure}[H]
\begin{center}
\scalebox{0.75}{
\begin{tabular}{p{1.25in}p{0.9in}p{1.6in}}
  \centerline{$C$}
  &
  \centerline{$D$}
  &
  \centerline{$C \tensor D$}
  \\
\begin{tikzpicture}[baseline=(A),->,shorten >=1pt,auto,node distance=1.4cm,on grid,inner sep=0.02cm,semithick,bend angle=25]
\node[initial,initial text=,positive] (A) {};
\node[positive] (B) [below left=of A] {};
\node[positive] (C) [below right=of A] {};
\node[positive] (D) [below=of B] {};
\node[positive] (E) [below=of C] {};
\node[positive,accepting by arrow] (F) [below right=of D] {};
\path [every node/.style={font=\footnotesize}]
(A)
edge [swap] node {$\singleton{p}{q}{\Send{i}}$} (B)
edge node {$\singleton{p}{q}{\Send{j}}$} (C)
(B)
edge [swap] node {$\singleton{p}{s}{\Send{i}}$} (D)
(C)
edge node {$\singleton{p}{s}{\Send{j}}$} (E)
(D)
edge [swap] node {$\singleton{p}{r}{\Send{i}}$} (F)
(E)
edge node {$\singleton{p}{r}{\Send{j}}$} (F)
;
\end{tikzpicture}
&
\begin{tikzpicture}[baseline=(A),->,shorten >=1pt,auto,node distance=1.4cm,on grid,inner sep=0.02cm,semithick,bend angle=25]
\node[initial,initial text=,negative] (A) {};
\node[positive] (B) [below left=of A] {};
\node[positive] (C) [below right=of A] {};
\node[positive,accepting by arrow] (D) [below right=of B] {};
\path [every node/.style={font=\footnotesize}]
(A)
edge [swap] node {$\singleton{p}{q}{\Receive{i}}$} (B)
edge node {$\singleton{p}{q}{\Receive{j}}$} (C)
(B)
edge [swap] node {$\singleton{q}{r}{\Send{k}}$} (D)
(C)
edge node {$\singleton{s}{q}{\Receive{l}}$} (D)
;
\end{tikzpicture}
&
\begin{tikzpicture}[baseline=(A),->,shorten >=1pt,auto,node distance=1.4cm,on grid,inner sep=0.02cm,semithick,bend angle=25]
\node[initial,initial text=,positive] (A) {};
\node[positive] (B) [below left=of A] {};
\node[mixed] (C) [below right=of A] {};
\node[positive] (D) [below left=1.4cm and 0.6cm of B] {};
\node[positive] (E) [below right=1.4cm and 0.6cm of B] {};
\node[mixed] (F) [below left=1.4cm and 0.6cm of C] {};
\node[positive] (G) [below right=1.4cm and 0.6cm of C] {};
\node[positive] (H) [below=of D] {};
\node[positive] (I) [below=of E] {};
\node[positive] (J) [below=of F] {};
\node[positive] (K) [below=of G] {};
\node[positive,accepting by arrow] (L) [below right=1.4cm and 0.4cm of I] {};
\path [every node/.style={font=\footnotesize}]
(A)
edge [swap] node {$\tau$} (B)
edge node {$\tau$} (C)
(B)
edge [swap] node {$\singleton{p}{s}{\Send{i}}$} (D)
edge node {$\singleton{q}{r}{\Send{k}}$} (E)
(C)
edge [swap] node {$\singleton{p}{s}{\Send{j}}$} (F)
edge node {$\singleton{s}{q}{\Receive{l}}$} (G)
(D)
edge [swap] node {$\singleton{p}{r}{\Send{i}}$} (H)
edge [swap, pos=0.7] node {$\singleton{q}{r}{\Send{k}}$} (I)
(E)
edge [swap] node {$\singleton{p}{s}{\Send{i}}$} (I)
(F)
edge [swap] node {$\singleton{p}{r}{\Send{j}}$} (J)
edge [swap, pos=0.75] node {$\singleton{s}{q}{\Receive{l}}$} (K)
(G)
edge node {$\singleton{p}{s}{\Send{j}}$} (K)
(H)
edge [swap] node {$\singleton{q}{r}{\Send{k}}$} (L)
(I)
edge [swap, pos=0.2] node {$\singleton{p}{r}{\Send{i}}$} (L)
(J)
edge [pos=0.2] node {$\singleton{s}{q}{\Receive{l}}$} (L)
(K)
edge node {$\singleton{p}{r}{\Send{j}}$} (L)
;
\end{tikzpicture}
\end{tabular}
}
\end{center}
\caption{Composition of basic automata, distinguishing internal and external actions}
\label{fig:automata:composition}
\end{figure}

%% file: fig/automata/composition-incompatible.tex
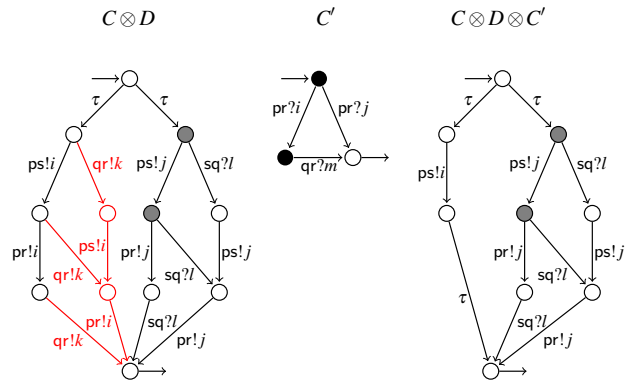
\begin{figure}[H]
\begin{center}
\scalebox{0.75}{
\begin{tabular}{p{1.65in}p{0.85in}p{1.15in}}
  \centerline{$C \tensor D$}
  &
  \centerline{\hspace{-0.5em}$C'$}
  &
  \centerline{$C \tensor D \tensor C'$}
  \\
\begin{tikzpicture}[baseline=(A),->,shorten >=1pt,auto,node distance=1.4cm,on grid,inner sep=0.02cm,semithick,bend angle=25]
\node[initial,initial text=,positive] (A) {};
\node[positive] (B) [below left=of A] {};
\node[mixed] (C) [below right=of A] {};
\node[positive] (D) [below left=1.4cm and 0.6cm of B] {};
\node[positive,red] (E) [below right=1.4cm and 0.6cm of B] {};
\node[mixed] (F) [below left=1.4cm and 0.6cm of C] {};
\node[positive] (G) [below right=1.4cm and 0.6cm of C] {};
\node[positive] (H) [below=of D] {};
\node[positive,red] (I) [below=of E] {};
\node[positive] (J) [below=of F] {};
\node[positive] (K) [below=of G] {};
\node[positive,accepting by arrow] (L) [below right=1.4cm and 0.4cm of I] {};
\path [every node/.style={font=\footnotesize}]
(A)
edge [swap] node {$\tau$} (B)
edge node {$\tau$} (C)
(B)
edge [swap] node {$\singleton{p}{s}{\Send{i}}$} (D)
edge [red] node {$\singleton{q}{r}{\Send{k}}$} (E)
(C)
edge [swap] node {$\singleton{p}{s}{\Send{j}}$} (F)
edge node {$\singleton{s}{q}{\Receive{l}}$} (G)
(D)
edge [swap] node {$\singleton{p}{r}{\Send{i}}$} (H)
edge [swap,pos=0.7,red] node {$\singleton{q}{r}{\Send{k}}$} (I)
(E)
edge [swap,red,pos=0.4] node {$\singleton{p}{s}{\Send{i}}$} (I)
(F)
edge [swap] node {$\singleton{p}{r}{\Send{j}}$} (J)
edge [swap,pos=0.7] node {$\singleton{s}{q}{\Receive{l}}$} (K)
(G)
edge node {$\singleton{p}{s}{\Send{j}}$} (K)
(H)
edge [swap,red] node {$\singleton{q}{r}{\Send{k}}$} (L)
(I)
edge [swap,pos=0.2,red] node {$\singleton{p}{r}{\Send{i}}$} (L)
(J)
edge [pos=0.2] node {$\singleton{s}{q}{\Receive{l}}$} (L)
(K)
edge node {$\singleton{p}{r}{\Send{j}}$} (L)
;
\end{tikzpicture}
&
\begin{tikzpicture}[baseline=(A),->,shorten >=1pt,auto,node distance=1.4cm,on grid,inner sep=0.02cm,semithick,bend angle=25]
\node[initial,initial text=,negative] (A) {};
\node[negative] (B) [below left=1.4cm and 0.6cm of A] {};
\node[positive, accepting by arrow] (C) [below right=1.4cm and 0.6cm of A] {};
\path [every node/.style={font=\footnotesize}]
(A)
edge [swap] node {$\singleton{p}{r}{\Receive{i}}$} (B)
edge node {$\singleton{p}{r}{\Receive{j}}$} (C)
(B)
edge [swap] node {$\singleton{q}{r}{\Receive{m}}$} (C)
;
\end{tikzpicture}
&
\begin{tikzpicture}[baseline=(A),->,shorten >=1pt,auto,node distance=1.4cm,on grid,inner sep=0.02cm,semithick,bend angle=25]
\node[initial,initial text=,positive] (A) {};
\node[positive] (B) [below left=of A] {};
\node[mixed] (C) [below right=of A] {};
\node[positive] (D) [below of=B] {};
\node[mixed] (F) [below left=1.4cm and 0.6cm of C] {};
\node[positive] (G) [below right=1.4cm and 0.6cm of C] {};
\node[positive] (J) [below=of F] {};
\node[positive] (K) [below=of G] {};
\node[positive,accepting by arrow] (L) [below left=1.4cm and 0.6cm of J] {};
\path [every node/.style={font=\footnotesize}]
(A)
edge [swap] node {$\tau$} (B)
edge node {$\tau$} (C)
(B)
edge [swap] node {$\singleton{p}{s}{\Send{i}}$} (D)
(C)
edge [swap] node {$\singleton{p}{s}{\Send{j}}$} (F)
edge node {$\singleton{s}{q}{\Receive{l}}$} (G)
(D)
edge [swap] node {$\tau$} (L)
(E)
(F)
edge [swap] node {$\singleton{p}{r}{\Send{j}}$} (J)
edge [swap,pos=0.7] node {$\singleton{s}{q}{\Receive{l}}$} (K)
(G)
edge node {$\singleton{p}{s}{\Send{j}}$} (K)
(I)
(J)
edge [pos=0.2] node {$\singleton{s}{q}{\Receive{l}}$} (L)
(K)
edge node {$\singleton{p}{r}{\Send{j}}$} (L)
;
\end{tikzpicture}
\end{tabular}
}
\end{center}
\caption{Incompatibility of $C \tensor D$ with $C'$, assuming $k \neq m$}
\label{fig:automata:composition-incompatible}
\end{figure}

%% file: sec/conclusion.tex
\section{Conclusion}

We proposed a language and tool design where simple concurrent objects
are checked exhaustively for compatibility and compliance, and more
complex objects can be partially verified via test cases which are
executed automatically. Although several non-trivial problems need to be
solved before this can applied to realistic examples, we hope to have
demonstrated the potential value of the approach and established some
connections with existing software development practices and design
patterns.

%% file: acknowledgements.tex
\paragraph{Acknowledgements.}

Supported by EPSRC (EP/K034413/1 and EP/L00058X/1) and COST Action
IC1201.